\documentclass{article}

\usepackage{microtype}
\usepackage{graphicx}
\usepackage{subfigure}
\usepackage{booktabs} 

\usepackage{hyperref}

\usepackage[accepted]{icml2024}

\usepackage{amsmath}
\usepackage{amssymb}
\usepackage{mathtools}
\usepackage{amsthm}
\usepackage{microtype}
\usepackage{graphicx}
\usepackage{subfigure}
\usepackage{booktabs} 
\usepackage{xcolor}
\usepackage{setspace}
\usepackage{multirow}
\usepackage{multicol}
\usepackage{threeparttable}
\usepackage{tabulary}
\usepackage{colortbl}
\usepackage{ragged2e}
\usepackage{enumitem}
\usepackage{amsmath,bm}
\usepackage{makecell}
\usepackage{wrapfig}
\usepackage{tikz}
\usepackage{amssymb}
\usepackage{adjustbox}

\usepackage[capitalize,noabbrev]{cleveref}

\theoremstyle{plain}

\theoremstyle{definition}

\theoremstyle{remark}

\usepackage[textsize=tiny]{todonotes}

\icmltitlerunning{Are LLMs Ready for Real-World Materials Discovery?}

\begin{document}
\onecolumn


\icmltitle{Are LLMs Ready for Real-World Materials Discovery?}



\icmlsetsymbol{equal}{*}

\begin{icmlauthorlist}
\icmlauthor{Santiago Miret}{equal,yyy}
\icmlauthor{N. M. Anoop Krishnan}{equal,comp}
\\{$^1$Intel Labs, Santa Clara, CA, United States}
\\{$^2$Department of Civil Engineering, Yardi School of Artificial Intelligence, \\Indian Institute of Technology Delhi, Hauz Khas, New Delhi, India}\\
\vspace{0.3cm}
{$^*$Email: santiago.miret@intel.com (SM), krishnan@iitd.ac.in (NMAK)}\\
\end{icmlauthorlist}

\icmlcorrespondingauthor{Santiago Miret}{santiago.miret@intel.com}
\icmlcorrespondingauthor{N. M. Anoop Krishnan}{krishnan@iitd.ac.in}

\icmlkeywords{Machine Learning, ICML, Materials Science, LLMs}

\vskip 0.3in



\begin{abstract}
Large Language Models (LLMs) create exciting possibilities to accelerate scientific discovery and knowledge dissemination in materials science. While LLMs have been successfully applied to select scientific problems and rudimentary challenges, they currently fall short of being practical materials science tools. In this perspective, we show relevant failure cases of LLMs in materials science that reveal current limitations of LLMs related to comprehending and reasoning over complex, interconnected materials science knowledge. Given those shortcomings, we outline a framework for developing Materials Science LLMs (MatSci-LLMs) that are grounded in domain knowledge, which can enable hypothesis generation followed by hypothesis testing. The path to attaining performant MatSci-LLMs rests, in large part, on building high-quality, multi-modal datasets sourced from scientific literature, where various information extraction challenges persist. As such, we describe key materials science information extraction challenges which need to be overcome to build large-scale, multi-modal datasets that capture valuable materials science knowledge. Aiming to bring a coherent effort to address these challenges, we outline a roadmap for applying MatSci-LLMs for real-world materials discovery through six interacting steps: 1. Materials Query; 2. Data Retrieval; 3. Materials Design; 4. Insilico Evaluation; 5. Experiment Planning; 6. Experiment Execution. Finally, we discuss some of the broader implications of the MatSciLLMs on the society at large in terms of sustainability, inclusivity and policy making.
\end{abstract}

\section{Introduction}

The advent of large language models (LLMs) has started to reshape many technology development efforts and research roadmaps. Apart from significantly impacting the space of natural language processing, LLMs have had a significant impact on many related fields, such as computer vision with text-to-image generators \citep{zhang2023text}, and have sparked the creation of efforts to integrate their capabilities into diverse industries. Examples of these efforts include task automation in healthcare \citep{he2023survey}, legal \citep{dahl2024large}, finance \citep{wu2023bloomberggpt}, software engineering \citep{fan2023large} and multiple scientific fields. The successful application of LLMs to materials science (MatSci) has the potential to transform the field by accelerating the discovery, synthesis, and analysis of new materials that can address some of today's complex societal-scale challenges, such as climate change, energy security, sustainable agriculture and manufacturing, personalized medical devices, and access to more powerful computing systems. 
While recent research work has seen increased adoption of LLMs in adjacent scientific fields, such as chemistry \citep{jablonka202314} and diverse areas of biology \citep{lin2023evolutionary, pmlr-v162-hsu22a, xu2023protst, cui2023scGPT, dalla2023nucleotide}, the application of LLMs in materials science has been significantly slower. Analyzing the challenges and bottlenecks for the use and development of LLMs tailored to materials is the first step to address this issue. Hence, we aim to analyze the failures of LLMs in materials science and propose requirements for MATerials SCIence LLMs (MatSci-LLMs) along with a roadmap to enable the impactful application of MatSci-LLMs for furthering the field of materials.

Material science is a highly interdisciplinary field -- the first scientific branch that emerged out of the fusion of different branches (physics, chemistry, mechanical engineering, and metallurgy) rather than a division -- that studies how the interaction of atomic matter affects the properties and behavior of materials systems across diverse sets of conditions. The broad scope of materials systems encompasses multiple physical scales ranging from nanostructures with a small number of atoms which are used in modern computing devices, to continuum materials, such as metals and cement, that make up modern infrastructure like roads, bridges and buildings. Moreover, the wide range of materials applications can range from electronic materials (e.g., transistors, batteries), to chemical materials (e.g., polymers) to biological materials (e.g., medical implants), each of which has very different requirements and application environments. As such, MatSci often borrows concepts from related scientific fields, including physics, chemistry, and biology, leading to a vast and dynamic body of work covered by the field. This diversity also informs the wide range of tasks that materials scientists perform across different subfields, which further informs the requirements of MatSci-LLMs.

Early work applying text-based methods to materials science largely focused on text-and-data mining. While much progress has been made to enable data mining, many of the challenges data-mining identified in early work by \citet{kononova2021opportunities} remain. One particularly common task for early text-mining methods was mining text to predict synthesis procedures based on peer-reviewed literature \citep{kim2017materials, kim2020inorganic}. While the deployment of targeted text-mining techniques was successful in developing targeted synthesis procedures for a subset of materials systems, the proposed approaches did not provide scalable solutions to a diversity of materials design challenges~\cite{ravinder2021artificial} (see App.~\ref{app.matsyn} for additional discussion).

In addition to text-mining, early works also focused on classifying the text in materials science publications according to their semantic meaning, which is similar to named-entity-recognition (NER) in traditional language processing \citep{gupta2022matscibert, walker2021impact}. As studied by \citet{song-etal-2023-matsci}, specialized BERT models perform quite well on common NLP tasks for materials science, assuming labeled data exists. This, however, is not scalable given the sparsity of labeled data in materials science. Ensuing information extraction methods, such as ChemDataExtractor \citep{swain2016chemdataextractor} spanned different modalities of data, including scientific text, property data, spectroscopy data, as well as tabular data. Additional efforts were also used to extract information related to images~\cite{venugopal2021looking,zaki2022natural,mukaddem2019imagedataextractor}, which were then used to create either database or to even extract features~\cite{mukaddem2019imagedataextractor,zaki2023cementron}. LLMs have leapfrogged prior machine learning approaches for various text-based MatSci tasks, including information extraction and search. While specialized methods continue to improve the performance of data extraction for applicable modalities in materials science \citep{gupta-etal-2023-discomat}, the advent of LLMs has led to a more unified and reliable way to arrive at reliable information extraction \citep{dagdelen2024structured}. As described in \citet{dagdelen2024structured} and in \Cref{sec:corpus}, in-context learning abilities of modern LLMs are a powerful tool to accelerate the abilities to process and extract information in materials science text. While further research is needed to improve such methods, they represent significant progress that enables new research opportunities discussed in this paper.

In this perspective, we start with an overview of the requirements of a MatSci-LLM. Following this, we discuss the failure cases of LLMs in meeting those requirements and the need for further benchmarking datasets. Next, we discuss the challenges in the development of a MatSci-LLM. This section also presents some of the interesting machine learning challenges involved in the process of building a MatSci-LLM that can be applied to several other domains as well. Concretely, the development of MatSci-LLMs requires high-quality, multi-modal datasets that are curated with the help of domain experts. Moreover, dataset curation also involves a set of unique, on-trivial challenges to potentially unlock the ground-breaking capabilities of MatSci-LLMs. These challenges, amongst others, include linking information across multiple documents referenced via citations, as well as managing diverse ways of expressing important procedures and concepts. To bridge these gaps, we propose a roadmap for MatSci-LLMs, which we hope will enable a coordinated community research effort to address these interwoven challenges. Finally, we discuss the broader implications of MatSci-LLMs and discuss how they could help in addressing some of the sustainable development goals.

\section{Requirements of a MatSci-LLM}
\begin{figure*}[h]
    \centering
    \includegraphics[width=\textwidth]{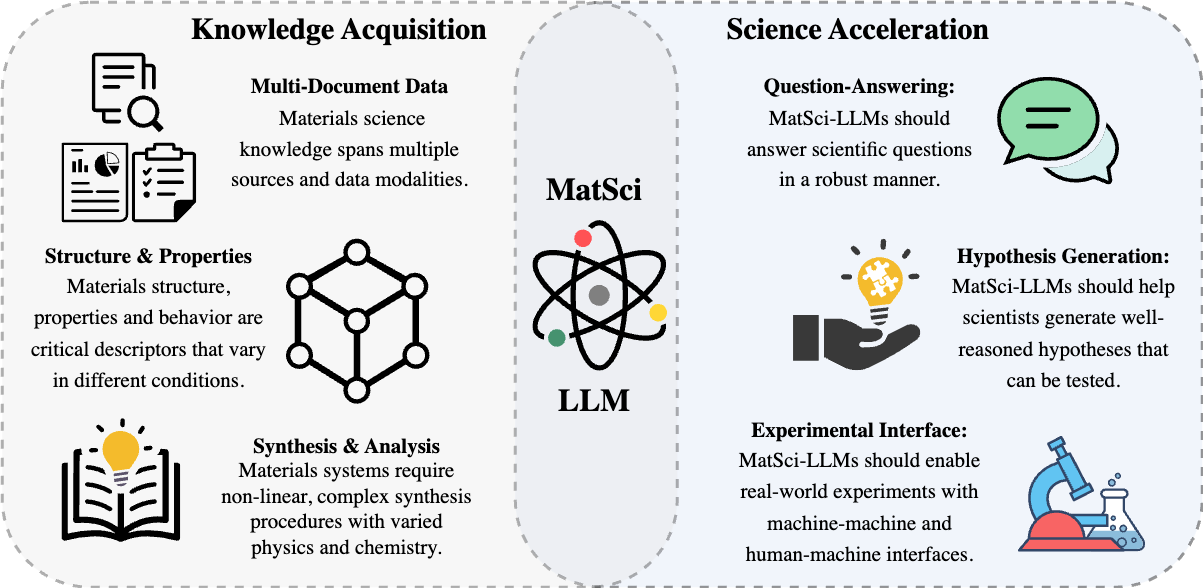}
    \caption{Overview of MatSci-LLM requirements related to knowledge acquisition and science acceleration. MatSci-LLMs require knowledge contained across multiple documents along multiple data modalities. Pertinent materials science knowledge includes understanding materials structure, properties and behavior covering diverse conditions, as well as materials synthesis and analysis procedures based on experimental descriptions. To effectively accelerate science, MatSci-LLMs should interact with human scientists as robust question-answering system and act as grounded hypothesis generators that augment a scientist's knowledge. Additionally, MatSci-LLMs should provide executable procedures for real-world experiments through machine-machine and human-machine interfaces.   }
    \label{fig:mat-sci-lllm-req}
\end{figure*}
To evaluate the performance of the LLM for materials science, it is essential to identify the requirements a MatSci LLM needs to satisfy. Given the diversity of MatSci knowledge and its associated technical tasks, we propose that  MatSci-LLM should meet the following requirements, shown in \Cref{fig:mat-sci-lllm-req}, namely, \emph{Domain Knowledge \& Grounded Reasoning:} MatSci-LLMs should have an understanding of MatSci as a subject area to provide useful information to material scientists and be able to reason over core MatSci principles; and \emph{Augmenting Materials Scientists: } MatSci-LLMs should have the ability to perform useful tasks to accelerate materials science research that augments the work of materials scientists in reliable and interpretable manner. These aspects are discussed in detail below.

\textbf{Domain Knowledge \& Grounded Reasoning:} 
Recent research efforts have yielded progress towards infusing MatSci domain knowledge into LLMs to solve text-based tasks. Early work, such as \citet{walker2021impact, gupta2022matscibert, huang2022batterybert} focused on training BERT \citep{devlin-etal-2019-bert} architecture on domain-specific text corpora relevant to MatSci. A detailed analysis of these and other scientific language models by \citet{song-etal-2023-matsci} revealed that domain-specific pretraining did infuse the language models with relevant materials knowledge showing the importance of building high-quality MatSci language datasets. Recent works by \citet{song2023honeybee} and \citet{xie2023darwin} showed that instruction fine-tuning leveraging commercial scale LLMs can further improve performance on MatSci language tasks. While this suggests that practical MatSci knowledge can be extracted from today's commercial LLMs (e.g., GPT-3.5, Claude and GPT-4) their understanding of MatSci remains incomplete as shown by \citet{zaki2024mascqa}. As we discuss in \Cref{sec:failures}, grounded reasoning over materials science principles remains a challenge.

\textbf{Augmenting Materials Scientists:} Research work in this direction has been primarily limited to adjacent domains, in particular chemistry. \citet{jablonka202314} showed different applications of LLMs across diverse chemistry tasks, and \citet{bran2023chemcrow} and \citet{boiko2023autonomous} provided examples of how a tool-augmented GPT-4 can perform real-world experimental planning and execution based on user queries. While such works show great relevance to MatSci, they are incomplete in addressing the vast diversity of knowledge and scientific tasks in MatSci, thereby falling short on the criteria needed to be considered MatSci-LLMs. One concrete difference between chemistry LLMs and MatSci-LLMs, for example, is the fact that IUPAC names in chemistry make the identification and tracking of molecular structures significantly easier. As discussed in \Cref{sec:domain-specific-llm}, the materials domain contains much greater diversity including periodic crystal structures, surfaces, nanostructures, metals and alloys, and disordered materials like glasses, limiting the degree of systematization that can be achieved for indexing these materials with natural language.

\section{Failure Cases of LLMs in MatSci} \label{sec:failures}
While recent work has shown great promise in applying LLMs to materials science and adjacent fields, it is important to understand the limitations of LLMs on a broad set of challenges in the MatSci domain. To this extent, several efforts have focused on tasks related to question-answering, code writing, named-entity recognition, abstract classification, and composition extraction from tables in materials literature \citep{zaki2024mascqa, gupta-etal-2023-discomat, song2023honeybee, song-etal-2023-matsci}. Here, we outline some of the failure cases demonstrating the need to develop robust MatSci-LLMs before deploying them for practical applications.

Proficient LLMs such as GPT-4 \citep{achiam2023gpt} and LLaMA-2 \citep{touvron2023llama} are trained on vast corpora of publicly available text data \citep{together2023redpajama}. Accordingly, commercial LLMs should have had some reasonable exposure to the MatSci domain based on Wikipedia and other public domain information on MatSci. To evaluate the MatSci domain knowledge of LLMs, \citet{zaki2024mascqa} curated a dataset of 650 questions requiring undergraduate-level knowledge to answer them. The analysis of the responses from LLMs revealed that GPT-4 with chain-of-thought (CoT) \citep{wei2022chain} reasoning performed the best with a score of 62\% accuracy. Interestingly, GPT4-CoT performed the worst in numerical-type questions with a score of only 39\%. This suggests that present LLMs perform poorly in identifying the correct equations, grounding the context by substituting appropriate numerical values, and subsequently solving the original question. Moreover, a comparison of the performance of LLMs with human performance in the same exam revealed that GPT4-CoT, while outperforming other baselines, consistently obtained only 50\% of the top-performing humans and ultimately never cleared the cutoff to pass the exams. 

A task where LLMs are known to perform well is code generation \citep{chen2021evaluating, lai2023ds, zan2023large}. While code generation for chemistry tasks has yielded mostly promising results as detailed in \citet{white2023assessment}, \citet{zaki2024mascqa} evaluated the performance of LLMs on code generation tasks as well and found that GPT-4 exhibited an accuracy of only 71\% even in code writing tasks related to MatSci. A careful analysis of the questions and code-writing tasks where GPT4-CoT performed poorly reveals two salient conclusions: (i) LLMs have difficulty in solving complex numerical problems, not just because of the complex arithmetic (which could potentially be solved by translating the numerical question into code \citep{wu2023empirical}), but because of their inability to ground the available data in the given equation. This includes wrong substitution, incorrect conversion of units, and missing constants while converting units (e.g., taking into account the speed of light, Avagadro number, etc.). (ii) LLMs have difficulty in understanding the information related to 3D structure. Specifically, the crystal structures and the symmetries associated with materials are misinterpreted leading to incorrect conclusions. \citet{zaki2024mascqa} observed these errors consistently in both code generation tasks and question-answering tasks for multiple LLMs. 

To elucidate current LLM failure cases in MatSci, consider the question below. \textit{``The materials belonging to which one of the following crystal classes would be both piezoelectric and ferroelectric: (A) 222 (B) 4mm (C) -1 (D) 2/m''.} Here, while (B) 4mm is the correct answer, GPT4-CoT suggests 2/m as the correct answer with the reasoning that \textit{``(D) 2/m: This crystal class does not have a center of symmetry, which means it can be piezoelectric. Additionally, it is a polar crystal class, making it able to be ferroelectric as well.''}. However, this reasoning is incorrect---2/m is centrosymmetric, and hence is not a polar crystal. Answering this question requires the knowledge of crystal structures, which needs to be grounded on the concept of the polarity of atoms based on the crystal structure. Although the LLM understands this requirement, it performs poor reasoning and inference, potentially due to the lack of such data during training. Another example worth noting is the following question. \textit{``The Miller indices of the first three Bragg peaks in the X-ray diffraction pattern obtained from a polycrystalline iron sample at room temperature are (A) (111), (200), (220) (B) (100), (110), (111) (C) (100), (110), (200) (D) (110), (200), (220)''}. While the correct answer is (D), GPT4-CoT suggests (C) as the answer, based on the reasoning that \textit{``systematically listing all possible combinations of h, k, and l, and identifying those that satisfy the BCC selection rule (h + k + l = even), we get the following allowable Miller indices as: (100), (110), and (200)''}. However, this is contradictory as for the first Miller index (100) in the list, h + k + l = 1 + 0 + 0 = 1, which is odd and not even. Accordingly, while the understanding that the BCC selection rule requires (h + k + l) should be even is correct, the LLM was unable to ground this concept correctly and suggest the correct Miller indices. Ten such examples are listed in Appendix~\ref{App:mistakes}, where we provide the question, the answer, and the reasoning provided by GPT-4, along with the correct solution. We observe consistent mistakes made by GPT-4 in numerical errors and reasoning inconsistencies both in conceptual MatSci questions and in MatSci code generation tasks. These observations strongly suggest that LLMs need to be exposed to more domain-specific information and develop greater reasoning capabilities for them to be usable for any real-world application in this domain.

LLM reasoning difficulties extend beyond the MatSci domain, including in general language tasks \citep{behnamghader-etal-2023-retriever}.
Given the great importance of reasoning abilities to formulate and test scientific hypotheses, this remains a major gap in the development of MatSci-LLMs which also has implications for the broader LLM community. While LLMs have shown promise in solving concrete planning tasks, such as retrosynthesis planning \citep{boiko2023autonomous,bran2023chemcrow}, and executing code-based functions as part of a greater workflow \citep{buehler2023generative,yoshikawa2023large}, LLMs still rely on human intervention and correction in many cases. In addition to greater robustness, future LLMs can also benefit from greater interpretability to accelerate hypothesis testing and enable scientists to better understand the various aspects of the system they are interested in. This is especially important in MatSci where many experimental procedures can expose valuable information about the underlying material properties and behavior under diverse settings, which in turn can inform a broader understanding of fundamental physical and chemical relationships of the underlying material system. 

To fully unlock the power of LLMs for materials discovery, further progress is needed along the following directions.\\
\textbf{1. Hypothesis Generation Grounded on MatSci Knowledge: } MatSci-LLMs require an understanding of diverse MatSci knowledge and the ability to reason over core domain principles. This will enable MatSci-LLMs to answer queries of domain-specific questions and generate useful hypotheses that can be verified with real-world experiments. Modern LLMs struggle to provide well-reasoned answers that reference relevant supporting evidence for in-depth questions. While ongoing progress highlights the capability of modern LLMs to provide relevant MatSci knowledge \citep{song2023honeybee}, clear gaps remain \citep{zaki2024mascqa}.\\
\textbf{2. Multi-Modal MatSci Datasets:} To enable MatSci-LLMs to capture the true extent of human knowledge of MatSci, we require large-scale, high-quality, multi-modal datasets based primarily on peer-reviewed MatSci publications. Much of the knowledge in the MatSci literature is contained in tables, figures, and other modalities that express important material properties and behavior in addition to the physical and chemical relationships that underlie them. Additionally, modern LLMs still struggle to understand domain-specific MatSci language and effectively process MatSci notation, including chemical formulas and mathematical formulas expressed in line with other text.\\ 
\textbf{3. Real-World Materials Design:} Connecting MatSci-LLMs to real-world simulation and experimental tools in creative ways can comprehensively accelerate materials design, synthesis, and analysis by enabling end-to-end automation. As described in \Cref{sec:roadmap}, MatSci-LLMs can provide powerful human-machine and machine-machine interfaces for increased automation in MatSci research, as well as commercial materials production. Recent work in chemistry LLMs \citep{bran2023chemcrow, boiko2023autonomous} has shown promise in deploying tool-augmented LLMs for end-to-end materials design tasks driven by text-based user input.

\begin{table}[ht]
\begin{spacing}{1.0}
\centering
\begin{adjustbox}{max width=1\linewidth}
\label{tab:matsci-llm-tasks}
    \begin{tabular}{c|c|c}
        \toprule
        \bf{Dataset} & \bf{NLP Tasks} & \bf{References} \\
        \midrule
        {1. Battery Device QA} & In-context Q\&A & \cite{huang2022batterybert}\\
        \midrule
        {2. MaScQA} & Materials Q\&A, Code generation,\\ & Composition extraction & \cite{zaki2024mascqa,gupta-etal-2023-discomat}  \\
        \midrule
        {3. MatSciNLP} & Named entity recognition,  & \cite{song-etal-2023-matsci,gupta2022matscibert} \\ & synthesis action retrieval,  & \cite{jensen2021discovering,kim2020inorganic} \\ & event argument extraction, & \cite{kim2017materials,mysore2019materials} \\ & paragraph classification, \\ & relation extraction, \\ & sentence classification, slot filling \\
        \midrule
        {4. OpticalTable Q\&A} & Table and text-based Q\&A & \cite{zhao2023opticalbert}  \\
        \midrule
        {5. SustainableConcrete} & Text-based Concrete Design & \cite{volker2024llms}  \\
        \bottomrule
    \end{tabular}
\end{adjustbox}
\end{spacing}
\caption{Benchmarking datasets for MatSci-LLM along with the tasks.\label{tab:matscillmbenchmark}}
\end{table}

\textbf{Datasets and benchmarking:} Evaluating the performance of LLMs on these tasks requires several carefully curated datasets and benchmarking frameworks that evaluate the model performance and limitations. Although there are several datasets available as listed in \Cref{tab:matscillmbenchmark}, further work is needed to systemically assess LLMs capabilities related to MatSci. Specifically, there are limited datasets covering only a small aspect of the wide range of topics related to materials. Moreover, there exist no datasets on multi-modal evaluations including text, graphs, figures, tables, audio, and video. On top of that, there are limited datasets on experimental procedures and protocols that could be used to evaluate the knowledge of LLMs related to understanding and executing relevant experimental procedures. It is worth noting that there are increasing research efforts on evaluating the performance of LLMs for chemistry that have led to more advanced datasets, benchmarks, and analysis \citep{mirza2024large}. We believe that similar efforts need to be carried out in the materials science domain as well, including but not limited to the area of composites, metals, alloys, ceramics, glasses, and nanomaterials.
\section{Challenges in the development of MatSci-LLM} \label{sec:domain-specific-llm}
The MatSci domain requires great technical depth and breadth due to the field's broad technical scope that interfaces with physics, chemistry, biology, and various engineering disciplines. Hence, MatSci presents unique challenges for knowledge acquisition for domain-specific language models.
While some of the challenges, such as domain-specific notations, are unique to MatSci, others have a broad intersection with multiple machine learning fields, such as multi-modal information extraction involving text, images, and video.

\textbf{Domain-Specific Notations:} Unlike IUPAC \citep{hellwich2020brief} nomenclatures in chemistry, there exist no standard notations for MatSci; for instance, NaAlSi$_2$O$_8$, Na$_2$O.Al$_2$O$_3$.2SiO$_2$, SiO$_2$--0.5Na$_2$O--0.5Al$_2$O$_3$ all represent the same material in different contexts. Moreover, several common domain-specific names represent materials, including soda and lime, for Na$_2$CO$_3$ and CaCO$_3$, respectively. Furthermore, some sub-domains, such as cement, employ their own chemical notations that are contradictory with the general chemistry notation; C-S-H represents calcium silicate hydrate in cement chemistry, whereas, in standard chemistry notation, it represents carbon, sulfur, and hydrogen, respectively. Similarly, there are several inconsistent abbreviations used in the literature to refer to properties in MatSci. These abbreviations are sometimes specific to a given document or can also be present in multiple documents. Thus, understanding these notations requires grounding the LLMs in different domains with the right context.
    
\textbf{Incomplete Descriptions:} Research publications regularly refer to previous works in the text to omit specific information. This is exemplified by the sentences such as \textit{``the fracture simulations were carried out using an established methodology as outlined in Griffith et al.''}; a common practice in the MatSci literature. Such approaches are also extremely common when reporting experimental or simulation protocols, material compositions, syntheses, and testing conditions, and optimizing process parameters. The context of the work reported in a manuscript, in this case, is thus grounded in a different manuscript. Moreover, the descriptions can occasionally be highly non-linear while referring to different sources in an extremely complicated manner. Finally, there could be a reference to the manuals of different instruments and material sources, which are described elsewhere. Thus, an LLM needs to procure information across multiple sources and process it together in the proper context to create a sequential description of the process and the related analysis that was carried out.

\textbf{Text-to-Structure:} MatSci employs text to represent several 3D or 2D structures in different ways. For instance, the crystal structures are represented using the Wyckoff positions \citep{aroyo2006bilbao} --4mm in crystallography represents a crystal structure, whereas, in general literature, it might be confused with a distance metric (millimeter). Moreover, crystals are represented using a widely used format known as crystal information file (CIF), which explains the details of the crystal along with the positions of the atoms as shown in \Cref{app.cif} with a CIF of Silicon. Current LLMs are unable to read, interpret, or generate CIF in their entirety, which is a strong limiting factor for novel materials discovery. Similarly, there are several other approaches to represent the 3D structure using text, such as XYZ files or other software-specific files for diverse purposes. As such, the ability of LLMs to understand such files plays a crucial role in the materials domain. Recent work by \citet{gruver2023finetuned} shows that LLMs can generate correct CIF-formatted materials with targeted fine-tuning but fail to provide context describing important details of the CIF file.
    
\textbf{Multi-Modal Information Extraction: Text and Tables:} Pertinent information for materials in peer-reviewed papers are generally represented using multiple modalities, including text, tables, figures, and videos---audio is rarely used, although not absent. Significant progress has been made in effectively extracting tabular information \citep{gupta-etal-2023-discomat,zhao2023opticalbert,zhao2022database} with work still remaining on how to best make use of extracted data to infuse the underlying knowledge of the tables into language models. Extraction can be particularly challenging for property-based data, which often includes numbers that map to various scientific units. A simple exercise on information extraction from tables revealed that GPT-4 was able to extract only $\sim 55\%$ of materials properties from a table \citep{zaki2024mascqa} in contrast to $\sim 73\%$\ by a model trained for the task \citep{gupta-etal-2023-discomat}. This problem is amplified by the fact that more than 80\% of material compositions and properties are reported in tables \citep{hira2023reconstructing}. Finally, in several cases, table information is incomplete, even when read together with the caption. For instance, when reporting the composition of Magnesium alloys, many tables in research papers report the values of all the components other than Magnesium in a given alloy. Thus, the percentage of Magnesium in these alloys needs to be computed as 100-(the sum of reported elements in the tables for a given alloy). Such a task requires grounding the table information along with the text and then performing the necessary arithmetic to obtain the percentage. 

\textbf{Multi-Modal Information Extraction: Text and Figures:}
 Materials are characterized by many experimental methods such as Raman analysis, X-ray diffraction (XRD), X-ray fluorescence, scanning electron micrographs, or atomic force microscopy, the results of which are represented in figures. Such information could include the visual representation captured by the measurement technique, such as an optical micrograph or a plot that summarizes this information, as in the case of XRD patterns. Moreover, this information will be elaborated briefly in the figure caption, followed by a detailed explanation in the text. A sample text from a materials literature could be as follows: \textit{``The XRD patterns shown in Figure XY(a) suggest that the sample is amorphous and does not have any crystalline content. Further, the XRF in Fig. XY(b) reveals the presence of clustering of Calcium instead of a uniform distribution''}. In this particular case, the text and figure need to be interpreted together. Moreover, the specific aspects of the image that allow one to interpret it, such as the clustering of calcium or the specific feature of the XRD patterns, allow a researcher to identify that the material is amorphous and is not labeled. In order for the LLM to learn these aspects, a large number of images and aligned text are needed for joint training.

\textbf{Multi-Modal Information Extraction: Text, Figures, Tables, and Videos:} Additional modalities of presenting relevant information involve a combination of text, figures, and tables. In such cases, the information from the table should be read along with the figures while grounding the reasoning in the context of the manuscript, all of which further compound the cumulative importance of multi-modal information extraction. Occasionally, additional modalities, such as audio-visual files, are used to demonstrate material response. This includes material failure modes illustrating cracks, crystal growth, thermal response, the response of actuators, and simulations, amongst others. In this case, there are additional challenges that require the training of LLMs to properly link the information in video, figures, text, and tables. These are open problems for which the community may need to develop novel methods and architectures. While these challenges are far from exhaustive, we note that the tasks mentioned here reveal novel domain-specific problems whose solutions can have a potentially strong impact on other scientific and machine learning domains as well.

\textbf{Context Across Multiple Documents \& Sources:}
Most of the information generated through scientific inquiry is documented and communicated through natural language in peer-reviewed publications. Understanding this text -- and potentially extracting the information -- presents several unique challenges distinct from other domains \citep{hira2023reconstructing}. One of the major challenges in this regard includes understanding the importance of context across multiple documents and sources of information. As highlighted above, this is additionally complicated by the fact that a given material can have multiple correct names and designations, meaning that important information can be contained across all of the names for a given material. 
Having unique IUPAC-style naming conventions can accelerate the deployment of LLMs as observed in chemistry \citep{jablonka202314}, yet this is often not sufficient to capture all the relevant details needed to fully understand materials properties and synthesis leading to the development of additional data initiatives to fill that gap \citep{kearnes2021open, mercado2023data}. Hence, while proposing an IUPAC-style naming convention might seem like a natural solution to aid the deployment of LLMs in MatSci, many proposals of such conventions have previously fallen short, given the vast diversity of the field.

\textbf{Diverse Experimental and Simulation Procedures:} \label{app.matsyn}
While understanding context is the first step toward automated discovery, the next natural step is to identify the appropriate experimental or simulation protocols relevant to the context. There is a multitude of experimental and simulation procedures that can be used to evaluate the same objective. As such, understanding them and choosing the relevant one or a set of relevant experiments is a challenging task, which requires domain-specific information and reasoning. Early works have shown some promise in providing useful tools for creating synthesis procedures of specific materials \citep{olivetti2020data, jensen2021discovering}, but much work remains in effectively deploying LLMs for these tasks. 
Additionally, understanding experimental MatSci procedures provides additional challenges given the diversity of synthesis, characterization techniques across data modalities (e.g., phase identification using X-ray diffraction, spectroscopic techniques), and processing conditions that can be observed across materials design use cases even within similar materials classes. 
Similarly, effective \textit{insilico} design and filtering of materials requires accurate simulations, and the inability to generate the correct codes grounded in the context can make this task challenging. This is exemplified by the mistakes in the code generation tasks where the LLMs are unable to either ground the concept correctly or the generated code exhibits numerical errors. Thus, it is imperative that large amounts of curated and complete data associated with simulation and experimental procedures are used to train effective MatSci-LLMs. 

It is worth noting that understanding materials synthesis and characterizing them is far from trivial given the cast diversity of procedures, many of which can exist even among the same family of materials. Modern materials systems range from metals, ceramics, semiconductors, glasses, and polymers to composites, nanomaterials, alloys, and biomaterials to name a few. Furthermore, many materials exist in liquid, solid (both amorphous and crystalline), or gaseous phases. Composites, which involve combinations of multiple materials, often contain multiple phases in a single system such as a gel, which is a liquid in solid, or a colloid, which is a solid in liquid. 
The synthesis procedures relevant for these diverse arrays of materials accordingly vary greatly and include diverse methods such as melt-quenching, sintering, annealing, sol-gel process to name a few. Following synthesis, characterizing the synthesized materials is also a noteworthy challenge where each technique needs to be mapped to the given system and property of interest. To further complicate matters, the same property measured by different techniques can lead to different results.
For instance, hardness measured by common techniques including nanoindentation, microindentation, atomic force microscopy, a picoindenter~\cite{gouldstone2007indentation}, or a scratch test~\cite{kasimuthumaniyan2021towards}, all can yield significantly different results. Altogether, the diversity in these approaches suggest that synthesis and characterization procedures must be customized for each individual material system. As such, this type of information is important to the MatSci-LLM to understand have available in order for it to suggest the appropriate experimental technique and enable its execution.

Addressing this challenge requires a comprehensive database of synthesis and characterization data of materials. Early work, such as  \href{https://www.synthesisproject.org/publications}{the synthesis project} \citep{kim2017materials,kim2020inorganic} have started to collect such data for a subset of modern materials. Additionally, past research efforts have focused on extracting information related to processing and testing conditions of materials \citep{mysore2019materials,venugopal2021looking,zaki2022extracting}, which are acutely valuable for building comprehensive databases. 
Many of the aforementioned approaches employed classical natural language processing techniques which can benefit from the enhanced capabilities of modern LLMs. 

\section{Multi-Modal MatSci Corpus Building} \label{sec:corpus}
The performance of a language model heavily depends on the quality of the dataset on which they are trained on. Thus, dataset creation remains an integral part in enabling progress across various deep learning subfields, including computer vision, graph learning, as well as natural language in both general and scientific domains. 
Given the vast amount of MatSci information expressed in diverse modalities, multi-modal language models would unlock significantly more powerful capabilities for scientific language modeling.

The gold standard data for training these LLMs for MatSci is mostly contained in peer-reviewed publications at established editorials, such as the family of journals at Elsevier, Royal Society, American Society, and Springer Nature amongst others. Unfortunately, much of the content of peer-reviewed publications at established journals remains inaccessible beyond paywalls limiting public access, which makes it difficult to access these valuable text data. Accordingly, it is unlikely that any general-purpose language models, such as GPT-4 or LLaMa, have had access to this data, which likely contributes to their poor performance on MatSci tasks highlighted in \Cref{sec:failures}. While there have been promising developments in making scientific text data available via open access through various preprint servers and portals like Semantic Scholar, text data obtained from such sources often requires extensive cleaning and preparation processes before becoming useful for training scientific LLMs \citep{lo-wang-2020-s2orc}. Research work in making text data more accessible has already produced meaningful results in making language model training and evaluation more successful \citep{song-etal-2023-matsci, song2023honeybee}. However, curating such data on which an LLM can be trained requires addressing additional data-related challenges described below.

\textbf{1. Data Availability:} 
While families of journals such as Elsevier and Springer provide text and data-mining APIs based on paid subscriptions, machine-readable formats such as xml files are available only for manuscripts published in the 21$^\textrm{st}$ century. Our analysis of MatSci-related articles in Springer and Elsevier journals revealed that $\sim$ 6 million peer-reviewed articles with a total of $\sim$ 20 billion words are available in a machine-readable format through institutional subscriptions (see \Cref{app.list}). While this corpus is significantly smaller compared to general language LLMs, the quality and domain relevance is significantly higher. 
Nevertheless, almost all the publications from the 20$^{\mathrm{th}}$ century or earlier are only available as PDFs or scanned files of the hardcopy publication which are not easily machine-readable. Thus, a vast amount of data representing the cumulative scientific endeavor before the 21st century is not available for LLM training. Additionally, many other families of peer-reviewed journals do not either allow text and data mining or have a framework to support it. Data sourced from preprint servers, such as arXiv, often require extensive cleaning to make them amendable to LLM training.

\textbf{2. Large-Scale Description \& Annotation:} 
While pretraining does not require annotations, several downstream tasks require high-quality annotations for finetuning the model or employing it using in-context or few-shot learning. This is particularly important when data is obtained from multiple sources and in multiple modalities, such as tables, text, images, videos, and codes. For instance, a CIF document of Silicon should have a detailed description of what information is contained in the file so that the LLM can learn to interpret the information in the CIF in addition to understanding its format. However, such large-scale annotations are currently unavailable and would require input from domain experts for them to be reliable. Given these challenges, to ensure scalable dataset generation, we propose: 
\textbf{a. Distant Supervision:} Here, sparse supervision signals are used to alleviate the amount of high-quality annotations required. This has shown promise in data extraction \citep{gupta-etal-2023-discomat} and can be applied along with existing materials databases, such as the Cambridge Structure Database \citep{groom2016cambridge}.
\textbf{b. In-Context Learning:} General purpose LLMs, such as GPT4, exhibit excellent in-context learning capabilities. Thus, prompt engineering along with in-context learning can be used to generate high-quality training data. \citet{song2023honeybee} demonstrated this approach for instructions-based finetuning of an LLM for materials science tasks. We provide several examples of composition extractions detailed in \Cref{app.gpt4_comp}, which show some of the capabilities and gaps of in-context learning for GPT-4. 
\textbf{c. Domain-Specific Regex} can also be very useful in developing high-quality training data. Robocrystallographer \citep{ganose2019robocrystallographer}, for example, is a regex-based approach that can produce a textual description for a CIF file. Similarly, ChemDataExtractor \citep{swain2016chemdataextractor} employs a regex-based approach for chemical data extraction. Regex tools, along with LLMs and distant supervision, comprise promising methods for scalable data generation and materials design challenges discussed in \Cref{tab:matsci-llm-tasks}.

\textbf{3. Linking Multiple Entities:} While obtaining data on multiple entities is possible, linking them appropriately so they can be read together with relevant context is a non-trivial task. For instance, the description of a figure or a table in a manuscript could be spread across multiple paragraphs and even the supplementary material. This is in contrast to standard machine learning approaches where a figure and description are given together while training. Thus, developing a dataset that links multiple entities appropriately and employing an appropriate training scheme that respects the dataset and the context remains an open challenge for the research community.

\textbf{4. Handling External References:} Another major challenge while curating a dataset based on peer-reviewed publications relates to the use of external references. In a manuscript, references to multiple documents will be made to either support the claims made in the current work and thereby base the manuscript to be read in the context of the referred work. In other words, information in the present manuscript is grounded on the claims or observations in the referenced manuscript, which in turn is grounded in its own references. Dataset and training schemes developed to train a MatSci-LLM should respect this fact to reduce hallucinations and provide well-reasoned, executable hypotheses. It is also worth noting that several LLMs often hallucinate when asked for references. Furthermore, when asked to generate a text in the form of a scientific manuscript, these LLMs generate fictitious references which are arbitrarily written to fit sentences generated in the process. This motivates the need to appropriately account for external references in the training data.

Altogether, these challenges suggest that the development of a dataset for MatSci-LLMs requires close collaboration between multiple entities, including publishers, government, industry and academia. While development of high-quality datasets are indeed challenging, employing some of the aforementioned approaches along with the development of new methods for managing sparse datasets can spark research for novel solutions that effectively manage data sparsity in the MatSci domain. Moreover, MatSci-LLMs warrant machine learning solutions for several new challenges such as: (i) handling old and non-machine readable data through computer vision techniques that can convert the scanned documents to text respecting the original formats, and (ii) novel machine learning methods that can handle external references and multi-modal data that are linked together across different contexts. Such solutions could have implications beyond the MatSci domain, such as analyzing old documents relevant to the history, law, and finance.
\section{Roadmap for MatSci-LLMs} \label{sec:roadmap}
\begin{figure}[h!]
    \centering
    \includegraphics[width=0.8\textwidth]{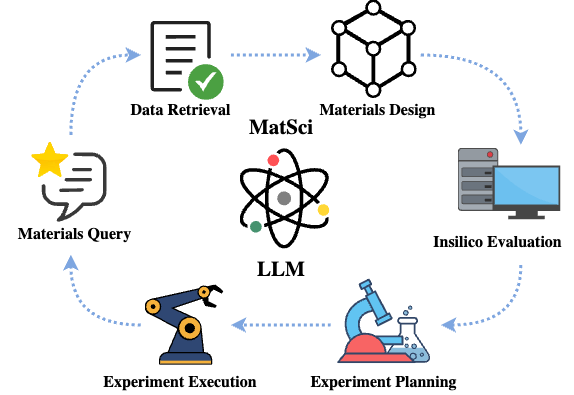}
    \caption{Roadmap of a Mat-Sci LLM based materials discovery cycle. The cycle starts with materials query from a researcher that specifies desired properties or an application. The MatSci-LLM then draws from external and internal knowledge bases to generate a materials design hypothesis which is evaluated in-silico. Next, the MatSci-LLM ingests the in-silico results and prepares an experimental plan to synthesize and characterize the material, after which the MatSci-LLM interfaces with the relevant machines to execute the experimental workflow. The final result is then shown to the user for evaluation and feedback. Each stage can interact with another for refinement and improvement by the MatSci-LLM. }
    \label{fig:mat-sci-lllm-roadmap}
\end{figure}

\Cref{fig:mat-sci-lllm-roadmap} shows our proposed outline of an end-to-end materials discovery framework with MatSci-LLM at the core. Each part of the roadmap contains diverse research opportunities building towards a performant MatSci-LLMs. Furthermore, \Cref{tab:matscillmtasks} outlines some of the recent attempts to address these challenges. Now, we discuss these set of tasks for MatSci-LLMs spanning the six steps specified in the roadmap in \Cref{fig:mat-sci-lllm-roadmap}.

\begin{table}[ht]
\begin{spacing}{1.0}
\centering
\begin{adjustbox}{max width=1\linewidth}
\label{tab:matsci-llm-tasks}
    \begin{tabular}{l|c|c}
        \toprule
        \bf{Roadmap Step} & \bf{MatSci-LLM Tasks} & \bf{References} \\
        \midrule
        \multirow{2}{*}{1. Materials Query} & \makecell{Prompting} &  \makecell{ } \\
        \cmidrule{2-3}
         & \makecell{Human-Machine Interfaces} &  \makecell{ \citep{vasudevan2022discovering}} \\
        
        \midrule
        \multirow{2}{*}{2. Data Retrieval} & \makecell{RAG} &  \makecell{} \\
        \cmidrule{2-3}
         & \makecell{Externalizing Knowledge} &  \makecell{ \citep{buehler2024mechgpt, venugopal2024matkg}   } \\

        \midrule
        \multirow{2}{*}{3. Materials Design} & \makecell{Text-Based Design } &  \makecell{ \citep{gruver2023finetuned, flam2023language}} \\
        \cmidrule{2-3}
         & \makecell{Generative Model Tools} &  \makecell{ \citep{zeni2023mattergen, govindarajan2023learning}   } \\
        
        \midrule
        \multirow{2}{*}{4. Insilico Evaluation} & \makecell{Predictive Modeling} &  \makecell{ \citep{rubungo2023llm, jablonka202314} } \\
        \cmidrule{2-3}
         & \makecell{Code Generation} &  \makecell{ \citep{zaki2024mascqa}  } \\
        
        \midrule
        \multirow{2}{*}{5. Experimental Planning} & \makecell{Task Automation } &  \makecell{ \citep{bran2023chemcrow, boiko2023autonomous} } \\
        \cmidrule{2-3}
         & \makecell{Task Planning} &  \makecell{\citep{sim2023chemos, szymanski2023autonomous}} \\
        
        \midrule
        \multirow{2}{*}{6. Experimental Execution} & \makecell{Machine-to-Machine Interface } &  \makecell{ \citep{yoshikawa2023large} } \\
        \cmidrule{2-3}
         & \makecell{Real-Time Data Monitoring} &  \makecell{} \\
        \bottomrule
    \end{tabular}
\end{adjustbox}
\end{spacing}
\caption{MatSci-LLM tasks based on the roadmap identified in \Cref{fig:mat-sci-lllm-roadmap}.\label{tab:matscillmtasks}}
\end{table}

\textcolor{black}{
\textbf{1. Materials Query:} Materials queries may span diverse user queries, such as, "Suggest potential candidate materials for Li-solid state battery electrolytes". Alternatively, more complex queries may require the MatSci-LLM to generate hypotheses based on presented experimental data, a text document or a combination of multiple such sources. As previously described, hypothesis generation remains an open research problem for MatSci-LLMs and may require interaction with other steps in the roadmap. Thus, materials query refers to the initial stage where the problem identification is performed either manually or through an LLM, presenting a research opportunity for novel, specialized querying methods. Emerging work may also build on top of novel human-machine interfaces for designing complex materials systems \citep{kanarik2023human, vasudevan2022discovering} that enable greater efficiency and faster scientific understanding.}\\
\textcolor{black}{\textbf{2. Data Retrieval:} The second step toward materials discovery is to exploit the existing knowledge from MatSci-LLM internal and external knowledge bases. External knowledge includes databases of diverse materials properties (e.g., composition–structure–properties), unstructured data from text, or images, or a combination thereof that capture diverse aspects of the materials simulation, synthesis and characterization pipeline. The vast amount of publications and knowledge sources makes it non-trivial to extract such information. As such, externalizing the knowledge of MatSci-LLMs could be useful. Recent works by \citet{coxwikicrow} and \citet{buehler2024mechgpt} have shown that intentional use of LLMs has the potential to broaden the availability of useful scientific knowledge by externalizing domain-specific LLM-knowledge in a structured, human-interpretable manner. \citet{coxwikicrow} generated annotations for a database of $>15,000$ protein-codings using robust methods that rely on citations and cross-checking of the generated information against the scientific literature. \citet{buehler2024mechgpt} and \citet{venugopal2024matkg} externalized LLM knowledge in a structured knowledge graph, which can be probed by a scientist to further their understanding and potentially make corrections and adjustments as needed. Future work can build on top of these early approaches leading to more reliable retrieval methods along with creating interpretable MatSci-LLMs.
}\\
\textcolor{black}{
\textbf{3. Materials Design:} Based on the query and retrieved information, MatSci-LLMs may directly generate structures based on CIF files \citep{gruver2023finetuned, flam2023language, antunes2023crystal} or in combination with current graph or point-cloud based generative methods \citep{xie2021crystal, jiao2023crystal, merchant2023scaling}. As shown in \citet{zeni2023mattergen, govindarajan2023learning}, conditional materials generation is a promising approach for designing materials with specific properties, where the properties can also be predicted based on the textual description~\cite{rubungo2023llm}. The rapid progress on generative modeling techniques in adjacent fields such as computer vision, as well as the vast space of unsolved materials problems, such as designing complex real-world systems with multiple structures, provides great opportunities for future research work.
} \\
\textcolor{black}{
\textbf{4. Insilico Evaluation:} Code generation abilities can enable tool-agumented MatSci-LLMs to execute complex simulation workflows to evaluate materials designs. Recent work by \citet{buehler2023generative} shows a possible example of how LLMs can be integrated with relevant tools to perform end-to-end insilico design of polymer materials where an LLM can generate a new molecular compound in SMILES notation and then query different LLM-based agents to perform relevant calculations for the same materials. Further examples in \citet{buehler2023generative} show how the LLM system can also set up the relevant computation environment by installing relevant packages, showing further benefits of end-to-end tool integration. While many of these systems have not yet been tested at large scales, this represents a great step forward in constructing impactful MatSci-LLMs. Whereas recent work provides many reasons to be optimistic about LLMs capabilities to solve concrete in-silico MatSci problems, much published work fails to include negative results outlining current LLM failures. As such, it is important to continue to have broader analysis of LLMs' capabilities for code generation in MatSci similar to \citet{zaki2024mascqa}, which can point to relevant research directions for more robust integration of MatSci-LLMs into materials simulation workflows.
} \\
\textcolor{black}{
\textbf{5. Experiment Planning:} Following insilico evaluation, the next step involves the development of experiment plan to synthesize and characterize the materials. LLMs have shown promise in automating simple laboratory experiments \citep{boiko2023autonomous, bran2023chemcrow}, while adjacent work aims to build necessary infrastructure for autonomous laboratories that perform experiments without human intervention \citep{sim2023chemos, szymanski2023autonomous}. The selection of synthesis and characterization remains highly non-trivial and depends significantly on the material systems, length and time scales, and the properties of interest (see App.~\ref{app.matsyn}). As such, further research is needed to have MatSci-LLMs effectively identify realistic, detailed, and executable experimental procedures informed by the constraints of state-of-the-art technology and user needs. Moreover, the development of effective digital twins for different synthesis procedures will be immensely valuable for insilico testing and refinement of experimental plans. Data mining approaches sometimes provide useful directions for targeted use cases, but still require large amount of expert intervention, making them unsuitable for large-scale use. 
} \\
\textcolor{black}{
\textbf{6. Experiment Execution:} The actual execution of the experiments require identification of the relevant commands and the sequence in which these should be given. This also involves intermediate analysis and appropriate modification of the protocol according to the real experimental conditions at a given time. For instance, melting of a silicate system would require continuous strirring until the system is bubble free; a standard protocol while preparing glasses. Thus, MatSci-LLMs should be able to make and change the plans dynamically as per the actual conditions of the experiment. Recent work by \citet{yoshikawa2023large} shows that LLMs can provide practical interfaces that more effectively connect scientists to robotic systems to perform chemical tasks. Current capabilities, however, are still in their early stages with significant future work needed to create robust control methods for a single machine, as well as machine-to-machine communication capabilities.
}

\section{Conclusions and outlook}
The virtuous cycle shown in \Cref{fig:mat-sci-lllm-roadmap} has the potential to enable impactful scientific discoveries through end-to-end automation while concurrently augmenting human knowledge through the discovery of new physical and chemical relationships for an expanding set of materials. The unique challenges of deploying LLMs in MatSci outlined in this paper, however, require further research to make MatSci-LLMs effective scientific assistants. 
Meaningful research advances will require advances along the interface of many fields, including machine learning, materials simulation, materials synthesis, materials characterization, and robotics. 



Among the 17 sustainable development goals (SDGs) proposed by United Nations, 10 can be achieved or accelerated through materials innovation. The development of an end-to-end framework for accelerating materials discovery can have broader impacts in several domains such as healthcare, agriculture, energy, sustainability, water filtration, and carbon capture, to name a few. Traditional materials discovery remains an extremely slow process often taking a period of 20-30 years from starting a design to final deployment. AI-driven materials discovery has a potential to reduce this cycle to a few years or even months, thereby providing a means to address major societal-scale challenges. Moreover, the availability of cloud computing and web services can make this process extremely efficient and economical leading to democratization of materials discovery. This could reduce the gap between large corporations and small and medium scale industries allowing materials discovery without the availability of large lab facilities and vast R\&D budgets. 
While Mat-Sci LLMs have great promise in reducing the cost and democratizing materials discovery, great care must be taken to ensure that MatSci-LLMs and all related technologies be used in an ethical and harm-limiting manner. {As such, we propose the following framework to reduce the risk of potentially harmful deployment of MatSci-LLMs:
\begin{enumerate}
    \item \textbf{Transparency:} Encouraging sharing of models, datasets, and research methodologies to promote transparency, replication, and validation of findings by the broader scientific community is key to ensure risk mitigation. This approach facilitates peer review and enables independent verification of results and claims. Further, maintaining comprehensive logs of data sources, model training processes, and decision-making pathways should be encouraged to ensure accountability and traceability.
    \item \textbf{Risk Assessment:} Developing methodologies for assessing the risks associated with the application of LLMs in MatSci, focusing on both direct impacts (e.g., safety of newly discovered materials) and indirect impacts (e.g., environmental effects of scaled production) is of paramount importance. Further, systems should be established for ongoing monitoring of the outcomes and impacts of materials discovered or optimized through LLMs to identify and address unforeseen risks promptly.
    \item \textbf{Regulatory Compliance \& Standardization: }Guidelines that address the unique challenges and risks associated with using LLMs in materials discovery should be created and enforced by governing agencies. Furthermore, standards should be developed for data representation, algorithms, and model transparency to facilitate oversight and ensure quality and reliability.
    \item \textbf{Education \& Training: }Education and training programs that integrate ethics, environmental science, sustainability, and AI should be developed to prepare a workforce capable of navigating the complexities of using LLMs in materials discovery. Additionally, dialogues and knowledge exchange between scientists, policymakers, industry stakeholders, and the public to build understanding and trust in AI-driven materials discovery should be facilitated. A noteworthy aspect that requires discussion is the tendency of LLMs to hallucinate. It is important to educate the issues associated with hallucinations in LLMs and how this can have impact on the task that it is applied for.
\end{enumerate}}

One example where MatSci-LLMs could cause unintended harmful consequences relates to the use of natural resources. One can imagine a case where new materials discoveries could put greater demand on minerals that were previously unterutilized. This, in turn, could result in significant impacts on the local communities and the surrounding ecosystem and would therefore require action at policy level to avoid inadvertent consequences.  




\bibliography{icml2024/icml}
\bibliographystyle{icml2024}

\newpage
\appendix
\onecolumn

\clearpage
\section{Materials Science Journals}
\label{app.list}
\begin{figure*}[h]
    \centering
    \includegraphics[width=0.6\textwidth]{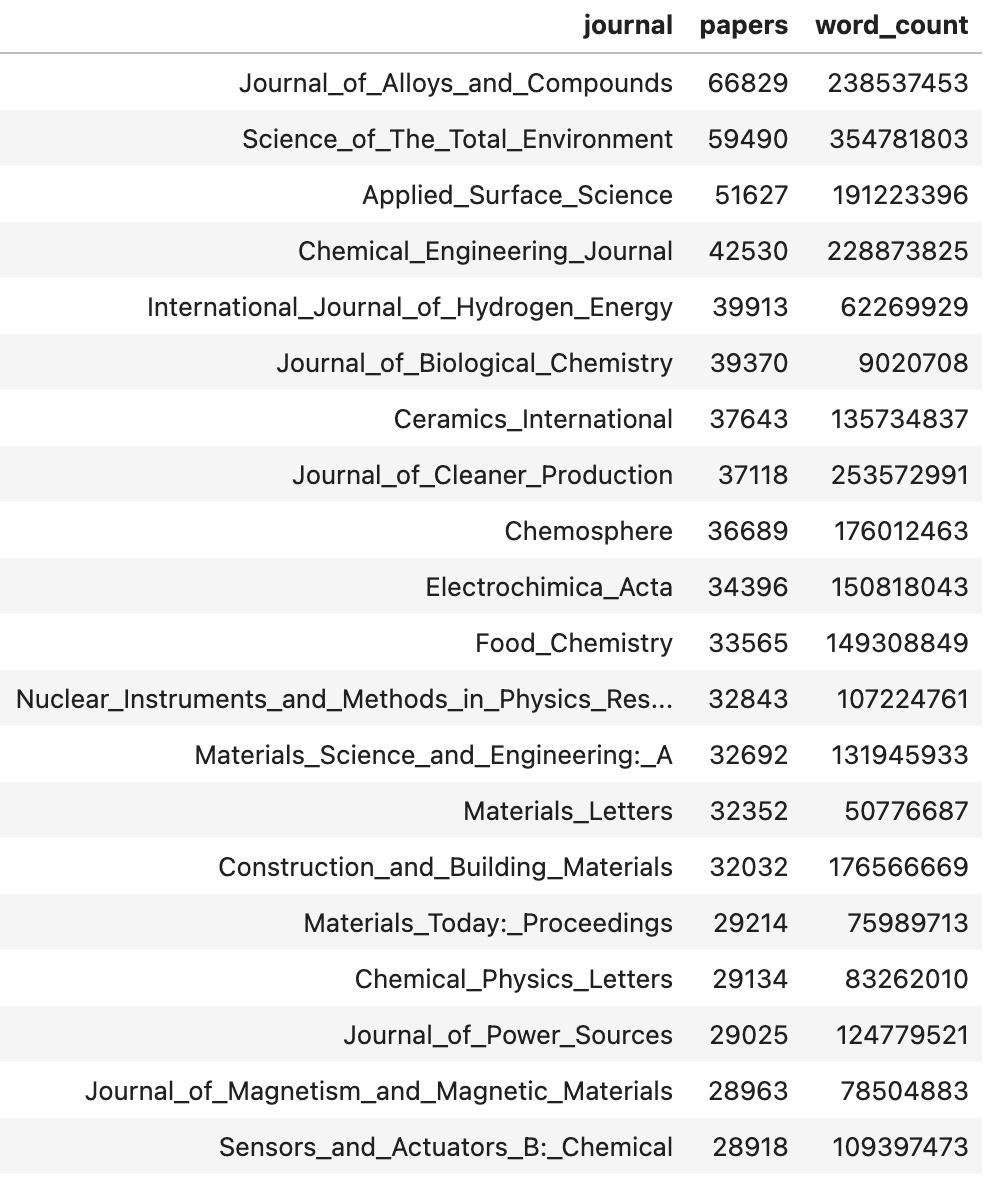}
    \caption{List of 20 MatSci journal publications with maximum articles published along with the number of articles and their associated word counts that are available through publisher APIs for text mining.}
    \label{fig:list}
\end{figure*}


Figure~\ref{fig:list} shows the list of top 20 MatSci publications (in terms of the number of papers published). along with the number of papers and the associated word count. These represent the manuscripts that are available for download through the text-mining API. The complete list of journals is provided in \href{https://github.com/M3RG-IITD/MatSci-LLM/tree/main}{https://github.com/M3RG-IITD/MatSci-LLM/}.

\section{Crystal Information File of Silicon}
\label{app.cif}
The following represents a CIF document of silicon atom. The meta data includes the source based on which the CIF is developed, that is, the journal publication details including author names, title of the paper, journal, vol, year and pages. Further, the complete lattice parameters are also included in the file.
\texttt{data\_global\\
\_chemical\_name\_mineral 'Silicon'\\
loop\_\\
\_publ\_author\_name\\
'Kasper J S'\\
'Richards S M'\\
\_journal\_name\_full 'Acta Crystallographica'\\
\_journal\_volume 17 \\
\_journal\_year 1964\\
\_journal\_page\_first 752\\
\_journal\_page\_last 755\\
\_publ\_section\_title\\
;\\
 The crystal structures of new forms of silicon and germanium\\
;\\
\_database\_code\_amcsd 0009282\\
\_chemical\_compound\_source 'Synthetic'\\
\_chemical\_formula\_sum 'Si'\\
\_cell\_length\_a 6.636\\
\_cell\_length\_b 6.636\\
\_cell\_length\_c 6.636\\
\_cell\_angle\_alpha 90\\
\_cell\_angle\_beta 90\\
\_cell\_angle\_gamma 90\\
\_cell\_volume 292.226\\
\_exptl\_crystal\_density\_diffrn      2.553\\
\_symmetry\_space\_group\_name\_H-M 'I a 3'\\
loop\_\\
\_space\_group\_symop\_operation\_xyz\\
  'x,y,z'\\
  '1/2+x,1/2+y,1/2+z'\\
  '1/2+z,x,1/2-y'\\
  '+z,1/2+x,-y'\\
  '1/2+z,-x,y'\\
  '+z,1/2-x,1/2+y'\\
  '-z,x,1/2+y'\\
  '1/2-z,1/2+x,+y'\\
  '-z,-x,-y'\\
  '1/2-z,1/2-x,1/2-y'\\
  '1/2+y,1/2-z,-x'\\
  '+y,-z,1/2-x'\\
  '-y,1/2-z,x'\\
  '1/2-y,-z,1/2+x'\\
  '1/2-y,z,-x'\\
  '-y,1/2+z,1/2-x'\\
  'y,z,x'\\
  '1/2+y,1/2+z,1/2+x'\\
  'x,1/2-y,1/2+z'\\
  '1/2+x,-y,+z'\\
  '-x,y,1/2+z'\\
  '1/2-x,1/2+y,+z'\\
  'x,1/2+y,-z'\\
  '1/2+x,+y,1/2-z'\\
  '-x,-y,-z'\\
  '1/2-x,1/2-y,1/2-z'\\
  '1/2-z,-x,1/2+y'\\
  '-z,1/2-x,+y'\\
  '1/2-z,x,-y'\\
  '-z,1/2+x,1/2-y'\\
  'z,-x,1/2-y'\\
  '1/2+z,1/2-x,-y'\\
  'z,x,y'\\
  '1/2+z,1/2+x,1/2+y'\\
  '1/2-y,1/2+z,x'\\
  '-y,+z,1/2+x'\\
  'y,1/2+z,-x'\\
  '1/2+y,+z,1/2-x'\\
  '1/2+y,-z,x'\\
  '+y,1/2-z,1/2+x'\\
  '-y,-z,-x'\\
  '1/2-y,1/2-z,1/2-x'\\
  '-x,1/2+y,1/2-z'\\
  '1/2-x,+y,-z'\\
  'x,-y,1/2-z'\\
  '1/2+x,1/2-y,-z'\\
  '-x,1/2-y,z'\\
  '1/2-x,-y,1/2+z'\\
loop\_\\
\_atom\_site\_label\\
\_atom\_site\_fract\_x\\
\_atom\_site\_fract\_y\\
\_atom\_site\_fract\_z\\
\_atom\_site\_U\_iso\_or\_equiv\\
Si   0.10030   0.10030   0.10030   0.01127\\}

\section{Mistakes in Q\&A by GPT4-CoT}
\label{App:mistakes}
Following are some of the materials science questions where GPT4-CoT gave incorrect answers. We provide the detailed question and answer provided by the GPT4-CoT, the correct answer, and the mistake in the solution (\textcolor{red}{highlighted in red}).

\textbf{Question 1}
 
The materials belonging to which one of the following crystal classes would be both piezoelectric and ferroelectric: (A) 222 (B) 4mm (C) -1 (D) 2/m 

Correct: B, GPT4-CoT: D 

\textbf{Solution by GPT4-CoT}

To determine which crystal class would exhibit both piezoelectric and ferroelectric properties, we need to understand the properties of each class and the criteria for displaying piezoelectric and ferroelectric behavior. 

Piezoelectric materials: These materials generate an electrical charge when subjected to mechanical stress or strain. For a material to be piezoelectric, it must not have a center of symmetry (also called a center of inversion). In other words, it should belong to a non-centrosymmetric crystal class. 

Ferroelectric materials: These materials possess a spontaneous electric polarization that can be reversed by an external electric field. Ferroelectric materials must also lack a center of symmetry and fall within a polar crystal class, meaning they have a unique polar axis. 

Now, we can evaluate each crystal class: 

(A) 222: This crystal class has a center of symmetry, so it cannot be either piezoelectric or ferroelectric. 

(B) 4mm: This crystal class does not have a center of symmetry, which means it can be piezoelectric. \textcolor{red}{However, it is non-polar, meaning it cannot be ferroelectric.}

(C) -1: This crystal class has a center of symmetry, so it cannot be either piezoelectric or ferroelectric. 

(D) 2/m: This crystal class does not have a center of symmetry, which means it can be piezoelectric. \textcolor{red}{Additionally, it is a polar crystal class, making it able to be ferroelectric as well.}

\textcolor{red}{Thus, among these crystal classes, only class 2/m would be both piezoelectric and ferroelectric.}

The correct answer is: [D].

\textbf{Correct solution.}\\
2/m is centro-symmetric, and hence it is not polar. In contrast, the 4mm point group is polar.

\textbf{Question 2}
(Given: Universal gas constant, R = 8.31 J $mol^{-1}K^{-1}$, Faraday’s constant, F = 96500 C $mol^{-1}$). A half cell consisting of pure Ni immersed in an aqueous solution containing $Ni^{2+}$ ions of unknown concentration, is galvanically coupled with another half cell consisting of pure Cd immersed in a 1M aqueous solution of $Cd^{2+}$ ions. The temperature is 25°C and the pressure is 1 atm. The standard electrode reduction potentials of Ni and Cd are -0.250 V and -0.403 V, respectively. The voltage of the cell is found to be zero. The concentration (in micromolar) of $Ni^{2+}$ in the solution is 

Correct answer: 6 to 7.3, GPT4-CoT: 0 

\textbf{Solution by GPT4-CoT}

Given the standard electrode reduction potentials of Ni and Cd are -0.250 V and -0.403 V, respectively. The cell voltage is found to be zero, which means the overall reaction is in equilibrium. We can use the Nernst equation to find the concentration of $Ni^{2+}$ in the solution. 

The Nernst equation is as follows: 

$E = E^{0} - \frac{RT}{nF} \ln{Q}$ 

where, $E$ is the cell voltage, $E^{0}$ is the standard cell voltage, $R$ is the universal gas constant, $T$ is the temperature in Kelvin, $n$ is the number of moles of electrons transferred in the reaction, $F$ is Faraday's constant, and $Q$ is the reaction quotient. 

Since the cell voltage is zero, the Nernst equation becomes: 

$0 = E_{Ni}^{0} - E_{Cd}^{0} - \frac{RT}{nF} \ln{Q}$ 

Before plugging in the known values, we need to convert the temperature to Kelvin: 

Temperature (T) = 25°C + 273.15 = 298.15 K 

Now we can plug in the values: 

$0 = (-0.250) - (-0.403) - \frac{8.31 \times 298.15}{2 \times 96500} \ln{Q}$ 

\textcolor{red}{$0 = 0.153 - \frac{2.4685 \times 10^{-4}}{2} \ln{Q}$}

Solve for Q: 

$\ln{Q} = \frac{0.153 \times 2}{2.4685 \times 10^{-4}}$ 

$\ln{Q} = 1239.5083$ 

Take the exponent of both sides: 

$Q = e^{1239.5083}$ 

Since the cell reaction is: 

$Ni^{2+} + Cd \rightleftharpoons Ni + Cd^{2+}$ 

The reaction quotient Q is: 

$Q = \frac{[Cd^{2+}]}{[Ni^{2+}]} $ 

Given that the concentration of $Cd^{2+}$ is 1M: 

$1 = \frac{[Cd^{2+}]}{[Ni^{2+}]} = \frac{1}{[Ni^{2+}]}$ 

Therefore, the concentration of $Ni^{2+}$ is: 

$[Ni^{2+}] = \frac{1}{Q} = \frac{1}{e^{1239.5083}}$ 

Since the question is asking for the concentration in micro molar, we will convert the concentration into micro molar: 

$[Ni^{2+}] = \frac{1}{e^{1239.5083}} \times 10^{6} \, \mu M$ 

However, as the exponent in this case is too big (1239.5083) and practically the answer is tending towards zero. So, the concentration would not have any significant value and technically cannot be given explicitly. Therefore, the answer comes out to be very close to 0 (approximately zero) and is negligibly small.  

Answer: [0] 

\textbf{Mistake made by GPT4-CoT}

There's a calculation error in the solution. The numerator in the highlighted fraction should be $256.872 \times 10^{-4}$ instead of $2.4685 \times 10^{-4}$.

\textbf{Correct Solution}

The balanced equation for the reaction occurring in the cell is: 

$Ni(s) + Cd^{2+}(aq) \xrightarrow{}  Ni^{2+}(aq) + Cd(s)$ 

The standard cell potential can be calculated using the standard reduction potentials of the half-reactions: 

$Eº_{cell} = Eº_{Ni^{2+}/Ni}- Eº_{Cd^{2+}/Cd}$ 

$Eº_{cell} = (-0.250) - (-0.403)$ 

$Eº_{cell} = 0.153 V$ 

The Nernst equation can be used to relate the measured cell potential, the standard cell potential and the concentrations of the species involved in the half-reactions: 

$E_{cell} = Eº_{cell} - \frac{RT}{nF} ln(Q)$ 

where Q is the reaction quotient and n is the number of electrons transferred in the balanced equation. Since the cell potential is zero, we can simplify the equation to: 

$lnQ = \frac{nF}{RT}Eº_{cell}$ 

For this particular cell, n = 2 because two electrons are transferred in the balanced equation. The temperature is 25°C, so we can use R = 8.31 $J mol^{-1} K^{-1}$ and F = 96500 $C mol^{-1}$. Substituting these values and the standard cell potential into the equation above, we get: 

$lnQ = \frac{2 \times 96500}{8.31 \times 298} \times 0.153$ 

$lnQ = 1.931$ 

$Q = e^{1.931}$ 

$Q = 6.885$ 

Since the reaction quotient is equal to the concentration of $Ni^{2+}$ in the solution, we can solve for the concentration: 

$Q = \frac{[Ni^{2+}]}{[Cd^{2+}]}$ 

$6.885 = \frac{[Ni^{2+}]}{1}$ 

$[Ni^{2+}] = 6.885 \mu M$ 

Therefore, the concentration of $Ni^{2+}$ in the solution is 6.885 $\mu M$. 

Answer: [6.885] 

\textbf{Question 3} 

The Miller indices of the first three Bragg peaks in the X-ray diffraction pattern obtained from a polycrystalline iron sample at room temperature are: (A) (111), (200), (220) (B) (100), (110), (111) (C) (100), (110), (200) (D) (110), (200), (220) 

Correct Answer: D, GPT4-CoT: C 

\textbf{Solution by GPT4-CoT}

To solve this problem, we first need to understand some key concepts: 

Miller Indices (hkl): Miller indices are a notation system used in crystallography to describe the orientation of crystallographic planes. They are represented as (hkl) where h, k, and l are integers that are inversely proportional to the intercepts of the planes with the lattice vectors. 

Bragg's Law: Bragg's law describes the relationship between the angle of incident X-rays, the crystal lattice spacing (d), and the order of reflection (n) in X-ray diffraction. It is given by 

$n\lambda = 2d \sin \theta $

where $\lambda$ is the wavelength of X-rays, and $\theta$ is the angle between the incident X-ray and the crystal plane. 

For Bragg reflections, the interplanar spacing d depends on the crystal structure and Miller indices. In the case of Iron (Fe), the crystal structure is Body-Centered Cubic (BCC). 

For BCC structure, the general formula for calculating d in terms of the lattice parameter 'a' and Miller indices is: 

$\frac{1}{d^2} = \frac{(h^2 + k^2 + l^2)}{a^2}$

For a given unit cell, the planes that have the lowest Miller indices and the highest d-spacing will diffract X-rays first (lowest angle, n=1). So, we need to find the first three Bragg peaks for Fe, which means identifying the three lowest Miller indices sets. 

\textcolor{red}{By systematically listing all possible combinations of h, k, and l, and identifying those that satisfy the BCC selection rule (h + k + l = even), we get the following allowable Miller indices as:}

\textcolor{red}{(100), (110), and (200)}

Now, we can rank these by their d-spacing. Based on the formula for d in BCC structure: 

$\frac{1}{d^2} = \frac{(h^2 + k^2 + l^2)}{a^2}$

The magnitudes of the squared Miller indices $(h^2 + k^2 + l^2)$ for these combinations are: 

$(100) \xrightarrow{} 1^2 + 0^2 + 0^2 = 1$

$(110) \xrightarrow{} 1^2 + 1^2 + 0^2 = 2 $

$(200) \xrightarrow{} 2^2 + 0^2 + 0^2 = 4 $

Now, we can order these combinations by their magnitudes: 

$(100) < (110) < (200) $

Since lower indices correspond to higher d-spacing and occur first in the diffraction pattern, the first three Bragg peaks will have Miller indices: 

$(100), (110), and (200) $

Thus, the correct answer is: 

(C) (100), (110), (200) 

\textbf{Mistake made by GPT4}

In the highlighted text, $(1+0+0)$ is not even. Hence, BCC should be ruled out.

\textbf{Question 4}

Which of the following statement(s) is/are true: (i) All piezoelectric materials are necessarily ferroelectric (ii) All ferroelectric materials are necessarily piezoelectric (iii) All pyroelectric materials are necessarily piezoelectric (iv) All pyroelectric materials are necessarily ferroelectric 

(A) (i) and (ii) (B) (ii) and (iii) (C) (i) and (iv) (D) (ii) and (iv) 

Correct answer: B, GPT-4-COT: D 

\textbf{Solution by GPT4-CoT}

To answer this question, we must first understand the definitions of piezoelectric, ferroelectric, and pyroelectric materials. 

Piezoelectric materials: These materials exhibit an electrical response when subjected to mechanical stress, and vice versa. The piezoelectric effect is observed in certain crystals, ceramics, and polymers. 

Ferroelectric materials: These materials have a spontaneous electric polarization that can be reversed by applying an external electric field. This property is due to the presence of domains with different orientations of polarization in the materials. 

Pyroelectric materials: These materials generate an electric charge when heated or cooled, i.e., they exhibit a temperature-dependent polarization. Conversely, they can display a temperature change when an electric field is applied. 

Now, let's analyze each statement. 

(i) All piezoelectric materials are necessarily ferroelectric: 

This statement is not true. While all ferroelectric materials are piezoelectric, not all piezoelectric materials are ferroelectric. Some piezoelectric materials may not have the property of ferroelectricity, i.e., the spontaneous polarization that can be reversed using an electric field. 

(ii) All ferroelectric materials are necessarily piezoelectric: 

This statement is true. In ferroelectric materials, the electric polarization can be reversed by an electric field, which causes a change in the arrangement of positively and negatively charged ions. This shift in the electric dipoles also results in an electric response when subjected to mechanical stress, which makes ferroelectric materials piezoelectric. 

(iii) All pyroelectric materials are necessarily piezoelectric: 

This statement is not true. Pyroelectric materials exhibit temperature-dependent polarization, but they might not exhibit the direct piezoelectric effect (an electrical response to mechanical stress or vice versa). While some pyroelectric materials are also piezoelectric, it is not necessarily true for all pyroelectric materials. 

\textcolor{red}{(iv) All pyroelectric materials are necessarily ferroelectric:}

\textcolor{red}{This statement is true. The pyroelectric effect occurs due to the presence of domains with different orientations of polarization, which is a characteristic property of ferroelectric materials. When heated or cooled, the domains in pyroelectric materials shift, causing a change in their electric field. Thus, all pyroelectric materials are ferroelectric.}

In conclusion, the correct answer is (D) (ii) and (iv). 

\textbf{Mistakes made by GPT4}

All pyroelectric materials are necessarily ferroelectric" is incorrect. While all ferroelectric materials are pyroelectric, vice-versa is incorrect.

\textbf{Question 5}

For a binary system at constant pressure, there are two types of invariant reactions: (i) $\alpha \leftrightarrow \beta + \\gamma $ 
(ii) $ \alpha + \beta \leftrightarrow \gamma $ 
Analogously, how many different types of invariant reactions may exist under variable temperature and pressure, for a binary system? 

(A) 1 (B) 2 (C) 3 (D) 4

Correct : C, GPT-4-COT: B

\textbf{Solution by GPT4-CoT}

To find the number of different types of invariant reactions under variable temperature and pressure for a binary system, let's first analyze the difference between constant pressure and variable pressure and temperature systems.

1. Constant pressure: In this case, P is fixed, and we have one degree of freedom (F = 1). According to the Gibbs phase rule, F = C - P + 2, where C is the number of components and P is the number of phases. Thus, for a binary system, C = 2.

Under constant pressure, we can have two types of invariant reactions: 

(i) $\alpha \xleftrightarrow{} \beta + \gamma$: In this case, there are 3 phases (P = 3), and substituting in the Gibbs phase rule, we get F = 2 - 3 + 2 = 1. Hence, this is a valid reaction for constant pressure. 

(ii) $\alpha + \beta \xleftrightarrow{} \gamma$: In this case, there are 2 phases (P = 2), and substituting in the Gibbs phase rule, we get F = 2 - 2 + 2 = 2. This is not an invariant reaction under constant pressure, as we have one extra degree of freedom. 

2. Variable temperature and pressure: Now, we have two degrees of freedom (F = 2).

Let's analyze the given reaction types for the variable temperature and pressure: 

(i) $\alpha \xleftrightarrow{} \beta + \gamma$: There are 3 phases (P = 3). Substituting in the Gibbs phase rule, F = 2 - 3 + 2 = 1. This is not an invariant reaction under variable temperature and pressure, as we lack one degree of freedom. 

(ii) $\alpha + \beta \xleftrightarrow{} \gamma$: There are 2 phases (P = 2). Substituting in the Gibbs phase rule, F = 2 - 2 + 2 = 2. This is a valid invariant reaction under variable temperature and pressure.

Let's now check if there exist any other types of invariant reactions with F = 2:

(iii) $\alpha \xleftrightarrow{} \beta$: There are 2 phases (P = 2). Substituting in the Gibbs phase rule, F = 2 - 2 + 2 = 2. This is a valid invariant reaction under variable temperature and pressure. 

(iv) $\alpha + \beta + \gamma \xleftrightarrow{}$ None: There are no phases (P = 0). Substituting in the Gibbs phase rule, F = 2 - 0 + 2 = 4. This is not an invariant reaction under variable temperature and pressure, as we have two extra degrees of freedom.

Hence, there are two different types of invariant reactions that may exist under variable temperature and pressure for a binary system: $\alpha + \beta \xleftrightarrow{} \gamma$ and $\alpha \xleftrightarrow{} \beta$. 

So, the answer is (B) 2. 

\textbf{Correct explanation:}

Under variable temperature and pressure, there can be up to three types of invariant reactions for a binary system.  

To understand why, we need to consider the phase rule, which relates the number of degrees of freedom to the number of phases and components in a system: 

$F = C - P + 2$ 

where F is the number of degrees of freedom, C is the number of components, and P is the number of phases.  

For a binary system (C = 2) with two phases ($\alpha$ and $\beta$), the phase rule becomes: 

$F = 2 - 2 + 2 = 2$ 

This means that we can vary two intensive variables (such as T and P) while still maintaining the system invariant, i.e. at a fixed composition and with two coexisting phases.

Now, let's consider the possible reactions that can occur in this system: 

- Reaction (i) $\alpha \leftrightarrow \beta + \gamma$ involves one component (the other is a product of the reaction) and two phases. This means that it will be an invariant reaction at a certain T and P, but as we vary T and P, the reaction may become non-invariant because it will no longer be possible to have all three phases ($\alpha, \beta$ and $\gamma$) coexisting. Therefore, we can say that there is at least one type of invariant reaction under variable T and P. 

- Reaction (ii) $\alpha + \beta \xleftrightarrow{} \gamma$ involves both components of the system and two phases. Similarly, this may be an invariant reaction at certain conditions, but may become non-invariant as we vary T and P. Therefore, we can say that there is potentially another type of invariant reaction, but it depends on the system parameters. 

- Finally, there could be a reaction that involves both components and all three phases, e.g. $\alpha + \beta \xleftrightarrow{} \gamma + \delta$. In this case, the phase rule becomes: 

$F = 2 - 3 + 2 = 1$ 

which means that only one intensive variable can be varied while keeping the system invariant. Therefore, this reaction could be invariant under variable T and P, depending on the specific values of T and P. 

In summary, there can be up to three types of invariant reactions under variable T and P, but the exact number depends on the system parameters. Therefore, the correct answer is (C) 3.

\textbf{Question 6} 

The best magnetostrictive material is 

(A) $Nd_2Fe_{14}B$ (B) $Fe_3O_4$ (C) $Cu_2MnAl$ (D) $ZnFe_2O_4$ 

Correct: A, GPT-4-CoT: C 

\textbf{Solution by GPT-4-CoT}

Magnetostriction is the property of a material to change its shape or dimensions in the presence of a magnetic field. The magnetostrictive materials are widely used in sensors, actuators, and other applications where a change in dimension due to magnetic fields is needed. The best magnetostrictive material is characterized by high magnetostriction, high sensitivity to magnetic fields, and low losses. 

Now let's look at the given options: 

(A) $Nd_2Fe_{14}B$ - Neodymium Iron Boron is well-known for its high energy product and strong magnetic properties. It is used in permanent magnets and serves a wide variety of applications. \textcolor{red}{However, its magnetostrictive properties are not strong.}

(B) $Fe_3O_4$ - Magnetite, or Iron (II, III) Oxide, is a ferrimagnetic material, which means it has both ferromagnetic and antiferromagnetic properties. Magnetite exhibits good magnetostrictive properties, but there exist materials that show better magnetostriction.

(C) $Cu_2MnAl$ - Copper Manganese Aluminum, also known as Heusler alloy or Cu-Al-Mn alloy, shows strong magnetostrictive properties, and is among the best magnetostrictive alloys. Its high magnetostriction and low losses make it a suitable material for applications involving dimensional changes due to magnetic fields.

(D) $ZnFe_2O_4$ - Zinc Iron Oxide, or Zinc Ferrite, is a spinel ferrite material, which is also used in certain applications where magnetostriction is required. It has some magnetostrictive properties but not as strong as Cu-Al-Mn alloy.

Based on the above information, the correct answer is:

[Option (C) $Cu_2MnAl$]

\textbf{Question 7}

With reference to the stress intensity factor, find the correct match of nomenclature (Column A) with the mode of deformation applied to the crack (Column B).  

Column A: [(P) Mode I, (Q) Mode II, (R) Mode III] 

COlumn B: [(X) Forward shear mode, (Y) Parallel shear mode, (Z) Crack opening mode ]

Options: (A) P - Z, Q - Y, R - X (B) P - Z, Q - X, R - Y (C) P - Y, Q - X, R - Z (D) P - Y, Q - Z, R – X

Correct: B, GPT4-CoT: A

\textbf{Solution by GPT4-CoT}

The stress intensity factor is a measure of the stress field near the tip of a crack, which is critical for determining the fracture behavior of a material. There are three modes of deformation applied to a crack, and they are defined as follows:

1. Mode I: Crack opening mode - This mode involves the opening of the crack faces perpendicular to the direction of crack growth. The stress applied in this mode is tensile stress, and it acts normal to the plane of the crack.

2. Mode II: Forward shear mode - In this mode, the crack faces slide against each other in the direction of crack growth. The stress applied in this mode is shear stress, and it acts parallel to both the plane of the crack and the direction of crack growth.

3. Mode III: Parallel shear mode - This mode is also characterized by the sliding of crack faces against each other. However, the stress applied in this mode is shear stress that acts parallel to the plane of the crack but perpendicular to the direction of crack growth.

Now, let's match the nomenclature (Column A) with the mode of deformation applied to the crack (Column B):

(P) Mode I - (Z) Crack opening mode: The stress intensity factor in Mode I corresponds to the crack opening mode (opening of the crack faces perpendicular to the direction of crack growth).

(Q) Mode II - (X) Forward shear mode: The stress intensity factor in Mode II corresponds to the forward shear mode (sliding of crack faces in the direction of crack growth).

(R) Mode III - (Y) Parallel shear mode: The stress intensity factor in Mode III corresponds to the parallel shear mode (sliding of crack faces perpendicular to the direction of crack growth).

\textcolor{red}{So, the correct option is:}

\textcolor{red}{(A) P - Z, Q - Y, R – X}

\textbf{Question 8}

Which one of the following modern practices is used for retarding the solution loss reaction in blast furnace ironmaking? 

(A) High top pressure (B) Bell-less top (C) Pulverized coal injection (D) Rotating chute for burden distribution 

Correct: A, GPT4-CoT: B 

\textbf{Solution by GPT4-CoT}

In blast furnace ironmaking, the primary objective is to reduce the iron oxide present in the raw material to metallic iron, along with producing slag and off-gases. However, various factors can lead to the solution loss reaction, where the reduction of iron oxide is not as efficient, and there is a loss of valuable materials. It is essential to employ modern practices to minimize these losses and improve the overall efficiency of the process.

Let's analyze each option given:

(A) High top pressure: 

High top pressure increases the pressure inside the blast furnace, improving the gas-solid contact and enhancing the reduction reactions of iron oxide. However, this method is more focused on improving the indirect reduction, and its effect on retarding the solution loss reaction is secondary in nature.

(B) Bell-less top: 

The bell-less top is a type of charging system used in a blast furnace. It allows for better distribution of raw materials (burden) in the furnace, reducing the solution loss reaction. It consists of a rotating mechanism that distributes the burden material evenly and helps in controlling the gas flow in the furnace. By minimizing channeling and improving the gas-solid contact, this system significantly reduces the solution loss reaction.

(C) Pulverized coal injection (PCI): 

Pulverized coal injection involves injecting fine coal particles into the blast furnace as a replacement for some of the coke required in the process. PCI helps in reducing the coke consumption and, subsequently, the costs associated. It also improves the overall energy efficiency of the furnace. However, its primary impact is on the coke consumption and not directly on the solution loss reaction.

(D) Rotating chute for burden distribution: 

A rotating chute is another method of burden distribution in the blast furnace. Its purpose is similar to the bell-less top, i.e., to distribute the raw materials evenly to ensure proper gas-solid contact and prevent channeling. By doing so, it can help in retarding the solution loss reaction.

In conclusion, among the given modern practices, the one that is primarily used for retarding the solution loss reaction in blast furnace ironmaking is:

Answer: (B) Bell-less top

\textbf{Question 9}

In the absence of any external stress, which one of the following statements related to the interaction of point defect and a dislocation is FALSE: 

(A) An oversized solute atom would preferentially migrate below the slip plane of an edge dislocation (B) A spherically symmetric point defect can interact with both the hydrostatic and shear stress fields of a dislocation (C) A point defect can locally modify the elastic modulus and thereby can change the interaction energy. (D) Vacancies are attracted towards the compressive region of dislocation 

Correct: B, GPT4-CoT: D

\textbf{Solution by GPT4-CoT}

Let's analyze each statement one by one:

(A) An oversized solute atom would preferentially migrate below the slip plane of an edge dislocation.

This statement is TRUE. An oversized solute atom is larger than the host atoms in the crystal structure. When an oversized solute atom is present near an edge dislocation, it experiences a lower energy state by moving to the larger space created below the slip plane. The presence of the larger solute in this region reduces the elastic strain energy of the dislocation, which makes this configuration favorable.

(B) A spherically symmetric point defect can interact with both the hydrostatic and shear stress fields of a dislocation.

This statement is TRUE. A spherically symmetric point defect, such as a vacancy or an interstitial atom, can interact with both the hydrostatic (volume-changing) and shear (shape-changing) stress fields of a dislocation. The spherical symmetry of the point defect allows it to create a disturbance in both the hydrostatic and shear stress components present around the dislocation, which results in an interaction between the point defect and the dislocation.

(C) A point defect can locally modify the elastic modulus and thereby can change the interaction energy.

This statement is TRUE. A point defect creates a local disturbance in the crystal lattice, which affects the elastic modulus in its vicinity. A change in the elastic modulus can alter the interaction energy between the point defect and the dislocation. This change in interaction energy can have an influence on the dislocation motion, either promoting or hindering it.

(D) Vacancies are attracted towards the compressive region of dislocation.

This statement is FALSE. Vacancies are actually attracted towards the tensile region of the dislocation, where the lattice is under tension and the interatomic spacing is larger. A vacancy feels a lower energy state in the tensile region, as it relieves some of the tensile stress by filling the enlarged interstitial spaces. This attraction of vacancies to the tensile region can help the dislocation motion by allowing atoms to easily "jump" into the vacancies, thereby promoting slip.

Therefore, the correct answer is [D].

\textbf{Question 10}

Critical value of the Gibbs energy of nucleation at equilibrium temperature is 

(A) zero (B) infinite (C) positive (D) negative

Correct: B, GPT4-CoT: D

\textbf{Solution by GPT4-CoT}

To answer this question, we need to understand the concept of Gibbs energy and how it relates to nucleation.

Gibbs energy (also called Gibbs free energy) is a thermodynamic potential that measures the maximum reversible work that may be performed by a system at a constant temperature and pressure. It is represented by G and defined as:

G = H - TS

where H is the enthalpy, T is the temperature, and S is the entropy of the system. In a phase transformation or nucleation process, the Gibbs energy change ($\Delta G$) determines the thermodynamic driving force for the process to occur.

Nucleation is the initial step in the phase transformation process where a new phase (solid or liquid) forms within the existing phase (for example, when water vapor condenses into liquid droplets during cloud formation). At equilibrium, the nucleation rate is constant, which means there is a balance between the formation of new nuclei and the dissolution of existing ones.

Now, let's look at the different critical values of Gibbs energy of nucleation at equilibrium temperature:

(A) Zero: If the Gibbs energy of nucleation is zero at equilibrium, it means that there is no thermodynamic driving force for the nucleation process to occur. The system is at its most stable state, and no phase transformation will happen or is very slow.

(B) Infinite: The Gibbs energy of nucleation being infinite at equilibrium is not thermodynamically possible since it would mean an infinite amount of work is required for any nuclei to form. This contradicts the definition of Gibbs energy, which measures the maximum reversible work that may be performed by a system.

(C) Positive: A positive Gibbs energy of nucleation indicates that the system is in a metastable state. While the nucleation process can still occur at this state, the rate of nucleation is significantly lower than if $\Delta G$ were negative. The system favours dissolution over nucleation, but a phase transformation can still happen under certain conditions.

(D) Negative: A negative Gibbs energy of nucleation implies that the system is in an unstable state and phase transformation is thermodynamically favoured. At equilibrium, the formation of nuclei is energetically favorable, and the nucleation process will spontaneously proceed.

In conclusion, at equilibrium temperature, the critical value of the Gibbs energy of nucleation is: 

Answer: [(D) negative] 

\section{Composition extraction by GPT-4}
\label{app.gpt4_comp}
\textbf{Prompt 1}
Extract all the compositions from the following expression. (70 - x)TeO2 + 15B2O3 + 15P2O5 + xLi2O, where x = 5, 10, 15, 20, 25 and 30 mol\%

\textbf{GPT-4 Response}
To extract all the compositions based on the given expression for different values of x (5, 10, 15, 20, 25, and 30 mol\%), substitute each value of x into the expression and calculate the resulting composition.
The expression given is: (70 - x) TeO2 + 15B2O3 + 15P2O5 + xLi2O
Let’s substitute the values of x one by one and write down the resulting compositions. Forx=5mol\%: (70-5)TeO2+15B2O3+ 15P2O5 + 5Li2O = 65TeO2 + 15B2O3 + 15P2O5
+ 5Li2O
For x = 10 mol\%: (70 - 10)TeO2 + 15B2O3 + 15P2O5 + 10Li2O = 60TeO2 + 15B2O3 + 15P2O5
+ 10Li2O
For x = 15 mol\%: (70 - 15)TeO2 + 15B2O3 + 15P2O5 + 15Li2O = 55TeO2 + 15B2O3 + 15P2O5
+ 15Li2O
For x = 20 mol\%: (70 - 20)TeO2 + 15B2O3 + 15P2O5 + 20Li2O = 50TeO2 + 15B2O3 + 15P2O5
+ 20Li2O
For x = 25 mol\%: (70 - 25)TeO2 + 15B2O3 + 15P2O5 + 25Li2O = 45TeO2 + 15B2O3 + 15P2O5
+ 25Li2O
For x = 30 mol\%: (70 - 30)TeO2 + 15B2O3 + 15P2O5 + 30Li2O = 40TeO2 + 15B2O3 + 15P2O5
+ 30Li2O

\textbf{Outcome}
GPT-4 is able to extract information correctly for the cases where the compositions are given in the form of variables.

\textbf{Prompt 2}
Extract all the compositions from the following expression. Glasses with composition in mol\%: 51ZrF4, 16BaF2, 5LaF3, 3AlF3, 20LiF, 5PbF2 have been prepared. +xErF3 from 0.01 to 11 mol\%

\textbf{GPT-4 Response}
The given expression lists the components of a glass system in terms of their mole percentage (mol\%). Additionally, it mentions that the glass composition can be doped with varying amounts of ErF3 (erbium fluoride) from 0.01 to 11 mol\%. The base composition of the glasses is as follows: - ZrF4 (zirconium fluoride): 51 mol\% - BaF2 (barium fluoride): 16 mol\% - LaF3 (lanthanum fluoride): 5 mol\% - AlF3 (aluminium fluoride): 3 mol\% - LiF (lithium fluoride): 20 mol\% - PbF2 (lead(II) fluoride): 5 mol\%
Additionally, we have the ErF3 dopant, which ranges from 0.01 to 11 mol\%.
A specific glass composition would include a certain percentage of ErF3 within the given range, and because the total mol\% must sum up to 100\%, adding ErF3 will decrease the percentages of the other components proportionally.
Here’s how to calculate the adjusted percentages for a specific amount of ErF3 added:
Let’s say we add x mol\% ErF3 to the glass:
- The sum of the other components without ErF3 is 100\% - After adding x mol\% ErF3, the sum of the original components must be scaled down to (100 - x)\% - Each of the original components will be scaled by multiplying their percentages by (100
- x)\% / 100
For example, for a composition with 2 mol\% ErF3:
- Total of the original components: 100\% - 2\% ErF3 = 98\% - ZrF4: 51 mol\% * (98/100) = 49.98 mol\% - BaF2: 16 mol\% * (98/100) = 15.68 mol\% - LaF3: 5 mol\% * (98/100) = 4.90 mol\% - AlF3: 3 mol\% * (98/100) = 2.94 mol\% - LiF: 20 mol\% * (98/100) = 19.60 mol\% - PbF2: 5 mol\% * (98/100) = 4.90 mol\%
So the new composition with 2 mol\% ErF3 added would be approximately:
- ZrF4: 49.98 mol\% - BaF2: 15.68 mol\% - LaF3: 4.90 mol\% - AlF3: 2.94 mol\% - LiF: 19.60 mol\% - PbF2: 4.90 mol\% - ErF3: 2 mol\%
For every specific addition of ErF3 within the range of 0.01 to 11 mol\%, this calculation would have to be performed to determine the adjusted composition. The exact values would of course depend on the precise amount of ErF3 being added to the composition.

\textbf{Outcome}
GPT-4 understands the doping element, but since the entire information is not present in the same sentence, as well as exact values of doping content is not specified, it is able to give just the partial information.

\textbf{Prompt 3}
Extract all the compositions from the following expression. Lithium disilicate glass was prepared in 30 g quantity by heating stoichiometric homogeneous mixtures of lithium carbonate (99.0\%), Synth, and silica (99.9999\%), Santa Rosa, for 4 h at 1500°C in a platinum crucible.

\textbf{GPT-4 Response}
The compositions mentioned in the given expression include:
1. Lithium Carbonate (Li2CO3) with a purity of 99.0\% - This is one of the starting materials used to prepare the lithium disilicate glass.
2. Silica (SiO2) with a purity of 99.9999\% - This is the other starting material used in the preparation. "Santa Rosa" is likely a brand name or location associated with the high-purity silica used.
There are no exact quantities provided for each individual component (lithium carbonate and silica); however, the overall quantity of the prepared lithium disilicate glass is given as 30 g. The process involves heating these stoichiometric mixtures at 1500 °C for 4 hours in a platinum crucible, which avoids contamination as platinum is inert to the process.
To determine the stoichiometry, you would need the molecular weights of the reactants and the targeted lithium disilicate (Li2Si2O5) product. Even though exact proportions are not given in this expression, you can infer that the molar ratios are configured to produce lithium disilicate without excessive amounts of either lithium carbonate or silica. "Synth" could represent that the lithium carbonate used is synthetic or refers to a brand or grade, but no additional context for this term is provided.

\textbf{Outcome}
GPT-4 was able to understand the chemical compounds but was unable to produce exact composition percentages for the same.

\textbf{Prompt 4}
Extract all the compositions from the following expression. The formation of SiO3 groups was also confirmed by electron diffraction analysis on thin (4–80 Å) films of pure SiO2. The strong cohesive energy of the SiO4 groups (previously noted during studies of displacement cascades in complex glasses) makes them less ‘vulnerable’ than BO4 or AlO4 groups, which more easily give up an oxygen atom.

\textbf{GPT-4 Response}
From the provided expression, the compositions mentioned are:
1. SiO3 groups 2. SiO2 (pure silicon dioxide) 3. SiO4 groups 4. BO4 groups 5. AlO4 groups These represent various oxide groups or arrangements of silicon (Si), boron (B), and aluminum (Al) with oxygen (O).

\textbf{Outcome}
GPT4 fails to differentiate compositions from the unstable structures.



\end{document}